\documentclass[letterpaper]{article}
\usepackage{aaai}
\usepackage{times}
\usepackage{helvet}
\usepackage{courier}
\usepackage{graphicx}
\usepackage{subfigure}
\usepackage{booktabs}
\begin{document}
%
\title{Race, Ethnicity and National Origin-based Discrimination in Social Media and Hate Crimes Across 100 U.S. Cities}
\author{Kunal Relia, Zhengyi Li, Stephanie H. Cook, Rumi Chunara\\
New York University\\
krelia, zl1499, sc5810, rumi.chunara@nyu.edu\\
}
\maketitle
\begin{abstract}
\begin{quote}
We study malicious online content via a specific type of hate speech: race, ethnicity and national-origin based discrimination in social media, alongside hate crimes motivated by those characteristics, in 100 cities across the United States. We develop a spatially-diverse training dataset and classification pipeline to delineate targeted and self-narration of discrimination on social media, accounting for language across geographies. Controlling for census parameters, we find that the proportion of discrimination that is targeted is associated with the number of hate crimes. Finally, we explore the linguistic features of discrimination Tweets in relation to hate crimes by city, features used by users who Tweet different amounts of discrimination, and features of discrimination compared to non-discrimination Tweets. Findings from this spatial study can inform future studies of how discrimination in physical and virtual worlds vary by place, or how physical and virtual world discrimination may synergize.  

\end{quote}
\end{abstract}

\section{Introduction}
Race, ethnicity or national-origin based discrimination (hereafter referred to as ``discrimination'') is a type of hate speech that systemically and unfairly assigns value based on race, ethnicity, or national-origin and affects the daily realities of many communities. Researchers have used a variety of proxy measures to assess discrimination at scale, such as policies \cite{kawachi2003neighborhoods}, or bias-motivated crimes \cite{sharkey2010acute}. However, policies usually have large spatial resolution, not all discrimination escalates to a crime, and as a measure, crimes don't describe any details regarding specific issues, antecedents or motivations of the crime that can be used to better illuminate and mitigate discrimination. This need for better understanding of discrimination is compounded, or brought to attention by recent increases in hate crimes in the United States (U.S.) \cite{VOAcrimesrise}. Simultaneously, a very compelling area of recent social media research is in the exploration of types of hate speech and potential links between physical events \cite{elsherief2018hate,elsherief2018peer,olteanu2018effect,muller2018making,muller2018fanning}. In particular, an examination of race, ethnicity and national-origin based discrimination on social media and it's spatial characteristics is warranted, given that hate crimes based on these motivations are the largest form of hate crimes in the United States \cite{FBI_HC_stat}.

Some prior social media research has focused on race-based discrimination experienced, and specifically on describing the racist concepts (e.g. appearance or accent related) most experienced by people of different races \cite{yang2018understanding}. For the United States specifically, research has shown that anti-Muslim hate crimes since Donald Trump's presidential campaign have been concentrated in counties with high Twitter usage (specific content was not parsed) \cite{muller2018making}. 
Further, some social media research has identified self-narration and targeted hate speech both as important \cite{yang2018understanding,elsherief2018hate}, but a gap remains to examine these linguistically and comparatively in their possible association with physical events. Self-narration is important to consider as 86\% of 18- to 29-year-olds have witnessed harassing behaviors online, and 60\% of those ages 30 and older, and 24\% of 18- to 29-year-olds have experienced mental or emotional stress as a result of their online harassment  \cite{duggan2014online}. Simultaneously, it is estimated there are approximately 10,000 uses per day of racist and ethnic slur terms in English on Twitter \cite{bartlett2014anti}. For targeted discrimination, research has examined the scope of targets (personal or towards groups) \cite{elsherief2018hate}. Finally, research hasn't examined how different forms of race, ethnicity and/or national origin-based discrimination on social media vary across the country, nor systematically assessed spatial differences in how people produce or discuss hate, or discrimination online.

We address these important gaps and build upon prior social media research by examining a specific type of hate speech: race, ethnic or national-origin based discrimination, enabling us to study these alongside hate crimes (as we can filter those by this same group of biases), across 100 different cities in the United States. This examination across the entire country allows us to assess the relationship across varying levels of urbanization, across different constituent properties of cities, and as well across different levels of social media usage in different places. While understanding the relationship between the two does not necessitate a causal pathway (nor do we aim to show one), this analysis helps to identify the way(s) in which social media may be relevant as a source for understanding structural discrimination, and helps illuminate how discrimination on social media may vary by place, in comparison to hate crimes. As well, our spatial analysis allows us to incorporate and assess linguistic differences associated with race-based discrimination on Twitter across cities in the United States. Specific contributions of this work are:

\begin{itemize}
\item Creation of a spatially-diverse training data set to account for local variations in race-based hate speech
\item Development of a multi-level classifier to automatically identify self-narration versus targeted \emph{race, ethnicity or national-origin based} discrimination on social media
\item Assessment of the relationship between social media measures of targeted and self-narration of race, ethnicity or national-origin based discrimination and hate crimes motivated by the same biases in 100 cities across the United States, controlling for demographic and other city-level attributes.
\end{itemize}


\section{Related Work}
\subsection{Social Media and Hate Speech}
There is a recent and growing literature in the social media research community on characteristics of hate speech. Beyond just detecting hate speech, research has gone further, for example, in analysis of the differences between personally-targeted and broadly-targeted online hate speech, showing linguistic and substantive differences \cite{elsherief2018hate}. Also, comparative study of hate speech instigators and targeted users on Twitter found personality differences in both, different from the general Twitter population \cite{elsherief2018peer}. The above work was focused comprehensively on any hate speech, which is defined as speech that attacks a person or group on the basis of attributes such as race, religion, ethnic origin, national origin, sex, disability, sexual orientation, or gender identity \cite{elsherief2018hate}. Given that in the research here, we want to describe the association between social media discrimination and hate crimes, we specifically narrow this research to \emph{race, ethnicity and national-origin based} hate speech, as hate crimes are described by different biases that motivate them, this being one of the categories (race, ethnicity and national-origin motivated hate crimes are often grouped so we could not separate these out for all of the years and cities considered, and moreover these are often grouped together in studies of the implications of discrimination \cite{williams2003racial}). In regards to race-specific hate speech, there is research distinguishing self-narration of racial discrimination and identifying which types of support are provided and valued in subsequent replies \cite{yang2018understanding}. A user-level analysis characterizing those who display hate speech on Twitter was performed by annotating users' entire profiles, showing differences in activity patterns, word usage as well as network structure \cite{ribeiro2018characterizing}.

\subsection{Comparing Online Hate Speech to Offline Events}
To-date there have been a few efforts in examining hate speech in social media to offline events. An analysis found that extremist violence leads to an increase online hate speech, using a counterfactual time series method to estimate the impact of the offline events on hate speech \cite{olteanu2018effect}. The social media data and crimes examined in this work were based on Arabs and Muslims specifically. Research by M{\"u}ller et al. showed that right-wing anti-refugee sentiment on Facebook predicts violent crimes against refugees in otherwise similar municipalities with higher social media usage. Essentially, in this work they compare how social media posts affect crimes within the same municipality compared to other locations in the same week. Though this link was found, this paper was focused in Germany, the content examined was manually collected from one Facebook group, and the social media data and crimes examined were based on anti-refugee sentiment specifically \cite{muller2018fanning}. In the United States, another study has shown that that the rise in anti-Muslim hate crimes since Donald Trump's presidential campaign has been concentrated in counties with high Twitter usage (specific content was not parsed) \cite{muller2018making}. In sum, related work on social media hate speech motivates that hate crimes are likely to have many fundamental drivers; social media can help illuminate some of these local differences (such as variation in xenophobic ideology or a higher salience of immigrants). To advance the validation of social media for this work, we focus on hate crimes biased by race, ethnicity and national-origin, and appropriately parse social media to understand the same type of discrimination. This enables us to compare discrimination in social media to offline events at scale across the United States.

\section{Data}

\subsection{Hate Crime Data}
\begin{figure}[t!]
\centering
\includegraphics[width=\columnwidth]{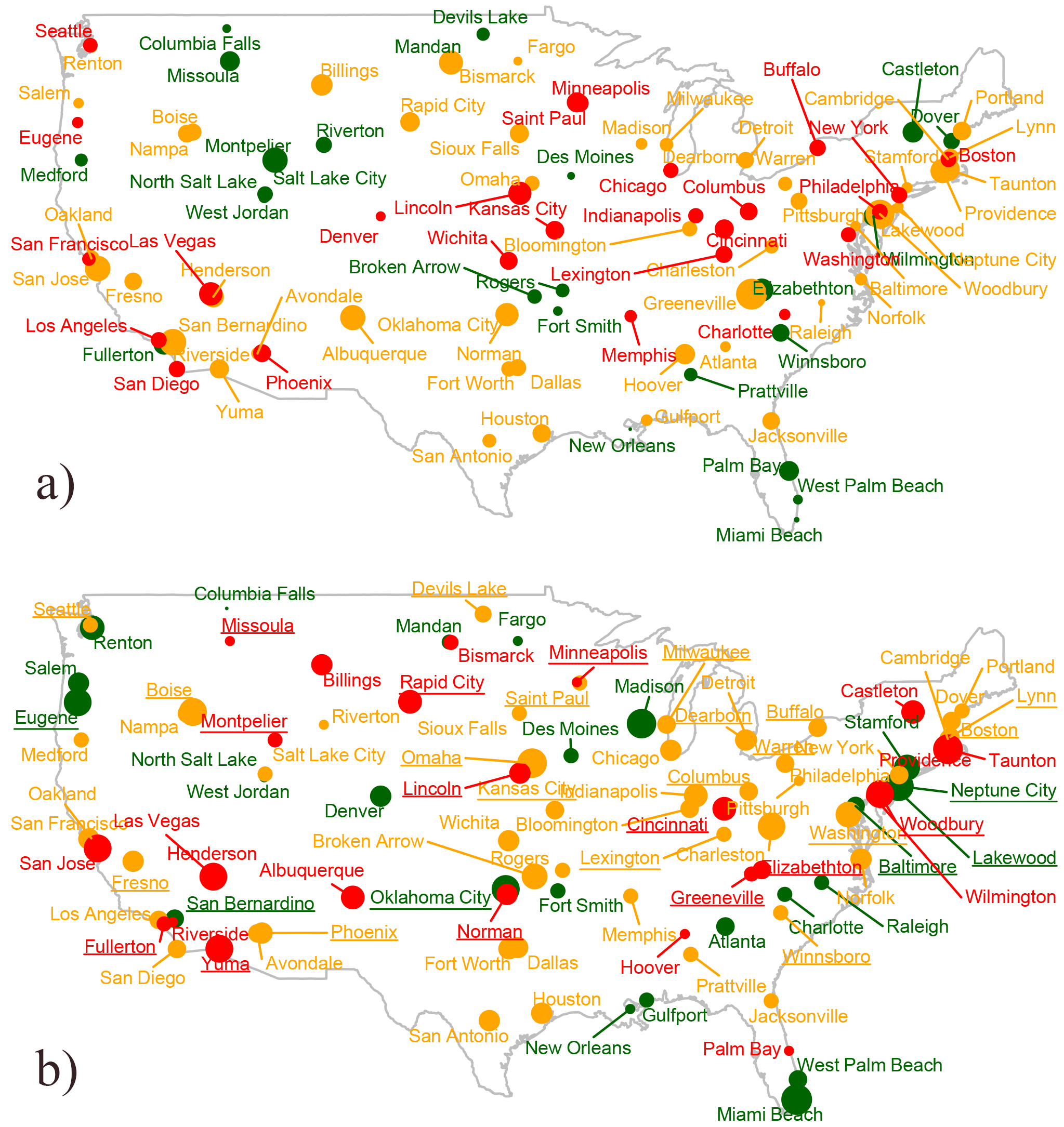}
\caption{U.S. map showing distribution of hate crimes and discrimination Tweets in the 100 cities. a) Color of labels assigned based on number of race, ethnicity and national-origin based hate crimes (green: lowest 25\%, yellow: between 25-75\%, red: top 25\%). Size of dots is based on the proportion of Tweets in that city that exhibit discrimination (self narration or targeted). b) Color of the labels is assigned based on proportion of Tweets that exhibit discrimination (self narration or targeted) (green: lowest 25\% of cities, yellow: 25-75\%, red: top 25\%). Size of dots is based on the ratio of number of discrimination Tweets to the number of unique users who produce them. Underlined cities have a targeted proportion of discrimination greater than half.}
\label{fig:maps}
\end{figure}

The Federal Bureau of Investigation (FBI) aggregates hate crime data under Congressional mandate. The biases that motivated the crimes are also recorded, broken down into specific categories (e.g. sexual-orientation or religious biases) \cite{FBI_HC_stat}. Agencies (generally metropolitans) of varying sizes contribute data. We used data from 2011--2016, which, at the time of analysis (9th November 2018), were the latest full years of data available, overlapping with our available Twitter data. Race, ethnicity and national-origin are combined as the motivating biases in some of the years, so for our study we focused on this entire group, aggregating hate crime data across these biases for years where they were delineated. One hundred cities which spanned a range of total number of hate crimes and locations were chosen. Cities in high and medium hate crime categories respectively were selected solely based on their hate crime numbers. Then, geographic regions which were under-represented in that group (e.g., Florida in the southeast, and Utah and Idaho in the midwest) were added in the low hate crime category to balance the geographical distribution of cities. Geographic distribution of the cities, shaded based on the number of race, ethnicity or national-origin based hate crimes by city is illustrated in Figure \ref{fig:maps}a. All 48 contiguous states, as well as Washington, D.C. are represented.

\subsection{Social media data}
We used Twitter's Streaming  Application Programming Interface (API) to procure a 1\% sample of Twitter’s public stream from January 1st, 2011 to December 31st, 2016. From this data set we selected Tweets made in the specified 100 cities using the ``place'' attribute. The place attribute contains the name of the city where a Tweet was made, determined using both the point and polygon coordinates associated with a Tweet. We manually accounted for changes in the way cities are described by name over time. Using place is computationally faster for matching city names as compared to mapping coordinates to cities using a polygon mapping algorithm with each Twitter JSON object. In total this resulted in 532 million Tweets. The text, time-stamp and location of the Tweets were used in discrimination classification; user id's were used in the bot analysis.


\subsection{Census data}
Census data for demographic and other city-attributes which, from domain knowledge may relate to discrimination, were included to control for their effect on the relationship between the online prevalence of discrimination and the number of race, ethnicity and national-origin based hate crimes. These included: the percentage of white, black, Asian, hispanic/latino, foreign born, female and ages 18-64 in the city. As well, population density (population per square mile) and median income (dollars) were included \cite{CensusData}. We used data from Census Quick Facts, as it combines statistics from the American Community Survey (ACS) with other surveys to give a broader view of a particular geography (includes population, density and income variables) \cite{CensusDataFullLink}. As well, Quick Facts uses ACS 5-year estimates which have increased statistical reliability compared with that of single-year estimates, particularly for small geographic areas and small population subgroups which we do have in this study. Further, the five-year estimates capture information across a large portion of our study years (2012-2016).

\section{Methods}
\subsection{Social Media Classification}
\begin{figure}[t!]
\centering
\includegraphics[width=\columnwidth]{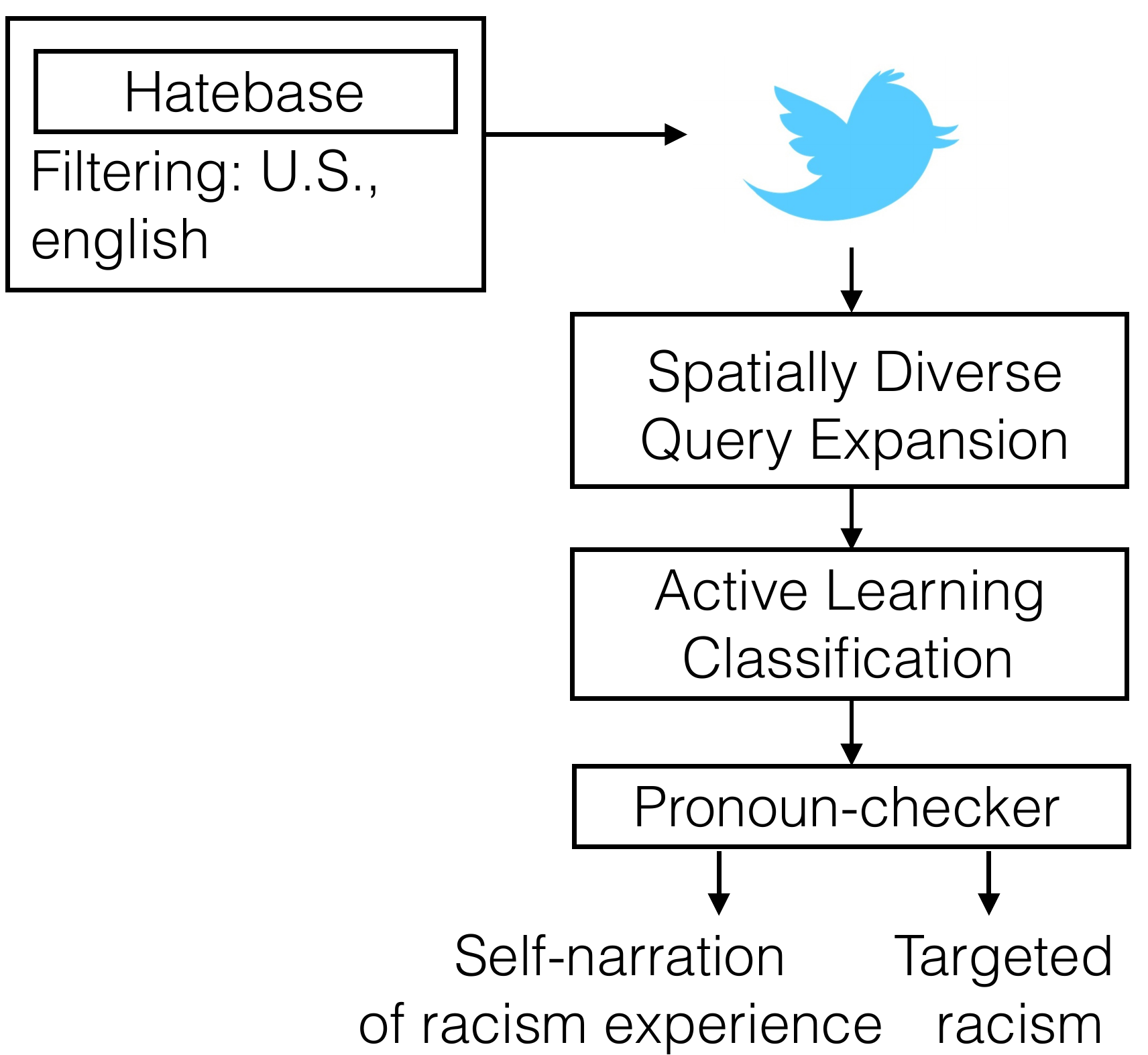}
\caption{Tweet processing pipeline.}
\label{fig:tweetclass}
\end{figure}

The classification pipeline to identify discrimination Tweets, and then delineate those into self-narration of discrimination or targeted is illustrated in Figure \ref{fig:tweetclass}. To classify Tweets we used shallow neural networks, which have shown improved performance over traditional classifiers \cite{dos2014deep,tang2014learning}, especially for short texts such as Twitter messages, which contain limited contextual information. The $n$-gram based approach and classifier parameters were the same as in previous work on discrimination classification that showed good performance \cite{relia2018socio}. As well, the overall approach followed from this same previous work in order to ensure that the resulting classified Tweets indicate \emph{discrimination} and not colloquial uses of keywords and phrases. For example, not all text that contains the ``n-word'' are motivated by racist attitudes \cite{relia2018socio}. More details are in the following sections.

\subsection{Spatially-diverse Training Data}
To improve classification performance, and account for possible language/terms specific to different locations across the United States, we developed a spatially diverse training data set. To do so, we used Tweets made in the top 11 cities ranked by total hate crimes. We specifically chose these cities, as there were more Tweets (245 million) made in these cities as compared to the next 39 ranked cities combined (206 million). As well, these cities provided geographic diversity, as we aim to capture language used in different parts of the country. 
In order to create a balanced training data that represents each of these 11 cities and each of the 6 years (2011 to 2016), we searched for hate speech keywords through a total of 73.42 million Tweets made in the United States in a randomly chosen week for each of the years from 2011 to 2016 . The list of keywords was selected from Hatebase.org \cite{elsherief2018peer,elsherief2018hate} such that each keyword had $>$10 sightings (a sighting is defined as ``actual incidents of hate speech for which we can establish both time and place'' \cite{Hatebase_sighting_definition}; and number of sightings is the number of times a term has occurred since March 25 2013, which is the beginning of the Hatebase project \cite{Hatebase_launch_blog}). We selected keywords in english only and removed keywords that were used more in non-derogatory contexts (e.g., Oreo and pancake), consistent with previous work \cite{elsherief2018hate,elsherief2018peer}. Out of the initial 73.42 million Tweets, 333,505 contained hate speech keywords, and out of which 21,490 were made in the selected 11 cities. Out of these 21,490 Tweets, 1,723 Tweets contained \emph{discrimination}-related keywords (ascertained by setting the nationality and ethnicity parameters on Hatebase to true). As the year 2016 had only 37 Tweets containing discrimination keywords made in the 11 cities, we added 200 Tweets containing discrimination keywords from the 11 cities from randomly chosen other weeks of 2016 so as to cover a wider range of keywords used in that year. Further, as Kansas City, MO didn't contain any Tweets containing discrimination keywords for 2011 and 2016, we collected more Tweets across all the six years from this city to have a temporally-balanced, better representation of it in the training dataset. In all, this resulted in 1988 Tweets to label.

\subsection{Labelling Data Procedure}
We used the services of Figure Eight\footnote{https://www.figure-eight.com/} (formerly known as Crowdflower) to label the training data, to capture and expand the keywords and phrases that are used in a discriminatory context. We clearly defined the criteria for labelling a Tweet (only the text of the Tweet was provided to the annotators) as indicating ``discrimination'' (versus ``no discrimination'') as \emph{a Tweet against a person, property, or society which is motivated, in whole or in part, by bias against race, ethnicity or national origin} to workers for annotation. Initial trial experiments and annotators’ review score on trial annotations confirmed the clarity of our instructions. As this project involved exposure of the annotators to potentially sensitive content, we clearly indicated the task is about Tweets that discuss discrimination, and created each Tweet as an individual task giving annotators a chance to discontinue at any point without losing payment if they felt uncomfortable. Each Tweet was labeled by at least two independent Figure Eight annotators, and all annotators were required to maintain at least an 80\% accuracy based on their performance on five test tasks. Annotators falling below this accuracy resulted in automatic removal from the task \cite{elsherief2018hate}. Out of the 1988 Tweets labeled, 1698 were labeled as discussing discrimination and the remaining as no discrimination. 

The label result was chosen based on the response with the greatest confidence of the labels. The confidence score (between 0 and 1) is calculated based on the level of agreement between multiple contributors, weighted by the contributors' trust scores \cite{FigureEight}. A high average confidence score of 0.92 resulted for the task, and our team manually labeled Tweets where there was a conflict of labels between the annotators. Finally, appending 14,012 Tweets not containing any discrimination keywords chosen equally from across the 11 cities and 6 years, resulted in a training data of 16,000 Tweets (1698 discrimination and 14,302 non-discrimination). This proportion is consistent with 10-13\% positive labels in a training dataset of 16,000 Tweets \cite{le2014distributed,dai2015document,elsherief2018peer}.


\subsection{Selection of Decision Boundary and Active learning for Classification}
For the shallow neural net classifier, selection of the decision boundary (threshold), 0.623, was made by optimizing the balance between precision and recall \cite{wulczyn2017ex}. The F1 score and AUC for this decision boundary, calculated by averaging the scores from a $k$-fold cross validation ($k$=10), were 0.85 and 0.89 respectively. Further, we used active learning to improve classification at the decision boundary. The entire active learning procedure involved first randomly sampling 10,000 Tweets from the top 11 cities, ranked by hate crimes, and classifying them using the shallow neural network. We then manually labelled the 1000 Tweets (5\% of Tweets on each side of the decision boundary) and appended these Tweets into the training data (active learning portion). We did this iteratively until performance of the classifier plateaued \cite{relia2018socio}. Finally, out of a total 17,000 Tweets in the training data, this resulted in 1987 discrimination and 15,013 non-discrimination Tweets. The average F1 score of the classifier improved to 0.86 and average AUC improved to 0.90 after this procedure.

\subsection{Delineating Targeted and Self-Narration Discrimination}
Previous social media research has discussed both \emph{targeted} \cite{elsherief2018hate} and \emph{self-narration} \cite{yang2018understanding} aspects of hate speech. Targeted discrimination is defined as someone being discriminatory as compared to self-narration of discrimination where someone is sharing their exposure to discrimination; either a direct experience or witnessing someone experience it. In line with existing work which used simple first-person pronoun filtering to identify self-narration of discrimination (racism) in Reddit posts \cite{yang2018understanding}, we used a similar first-person pronoun filtering approach. To categorize Tweets as self-narration, we selected Tweets that contain any of the first-person pronouns: \emph{I, me, mine, my, we, us, our, ours}. As simple first-person pronoun filtering resulted in a high false positive rate (e.g., \emph{``I think so-called white trash turns to white supremacy because, a., they're victimized by minority criminals, and b., they're uppity whites''}), we also required an absolute majority of number of first-person pronouns over number of second and third-person pronouns in a Tweet for a Tweet to be categorized as self-narration of discrimination. This condition helped to decrease the false positive rate by 75\%, measured by our team on a randomly chosen sample of 1000 Tweets. 

\subsection{Hate Crimes and Social Media Relationship}
To examine the relationship between race, ethnicity or national-origin based hate crimes and discrimination on social media by city, we use a regression modeling approach where the number of race, ethnicity or national-origin based hate crimes in all years, for each city, is the dependent variable, while accounting for other potential covariates. We caution that the approach does not imply that the social media measures directly causes hate crimes in different cities. Assessing this relationship would be a first step towards assessing any potential causal pathway, or the possible uses of social media discrimination in complement to hate crimes, for example to study issues at higher spatial-resolution than is available through hate crimes, such as at the neighborhood-level. Each model controlled for the census attributes discussed in the Census Data section. All analyses were conducted in R v3.5.2. 

\subsection{Linguistic Analysis}
To assess affect and linguistic-related features of discrimination on social media at the city, user and Tweet level, we used EMPATH. EMPATH is a tool that can generate and validate new lexical categories on demand from a small set of seed terms and capture aspects of affective expression, linguistic style, behavior, and psychological state of individuals from content shared on social media by deep learning a neural embedding across more than 1.8 billion words. The performance of EMPATH has been found to be similar to LIWC (considered a gold standard for lexical analysis), EMPATH is freely available, and EMPATH also provides a broader set of categories to choose from compared to LIWC \cite{fast2016empath}. We selected several affect-related features based on prior work in understanding self-narration of discrimination and discussion of racial equity \cite{yang2018understanding,de2016social}: \emph{positive emotion, negative emotion, disappointment, sadness, aggression, violence}. Motivated by studies regarding risk factors for racial/ethnic discrimination we also selected EMPATH features potentially related to socio-economic status (\emph{work, money}) and culture (\emph{night} e.g. nightlife) \cite{williams2003racial}. 

\subsubsection{City-level}
To understand the relationship between these linguistic features and hate crimes in a city, we used a regression modeling approach, where the number of race, ethnicity or national-origin based hate crimes in a city is the dependent variable, and the per-city normalized proportion of each linguistic feature are independent variables. Negative binomial regression is used to model the count data, and account for over-dispersion. We also present iterative model selection results (backwards step-wise model selection by exact Akaike information criterion) to protect against over-optimism and assess predictors that contribute a significant part of explained variance.

\subsubsection{User-level}
Here we examine the features used by users who discuss increased amounts of discrimination ($>$ 21 discrimination Tweets total) versus those who only had one discrimination Tweet. Those users with $>$ 21 discrimination Tweets represent those (0.01\% of all users with any discrimination Tweets) with the very highest number of discrimination Tweets in our dataset (more description of this range in the Descriptive Analyses of Discrimination Results section).

\subsubsection{Tweet-level}
At the Tweet level, we examined the distribution of all EMPATH linguistic features in discrimination versus non-discrimination Tweets. Using normalized means of the category counts for each group, we compute the correlation across these two types of Tweets and examine the odds of EMPATH feature categories likely to appear in discrimination Tweets.

\subsection{Discrimination Content by Bots}
Increasingly, there has been a recognition of bot accounts on Twitter that spread malware and unsolicited content that in particular has included public health and antagonistic content and towards eroding public consensus \cite{broniatowski2018weaponized}. Given the antagonistic nature of the content examined in this study, we also assessed and analyzed the prevalence of bots and bot-generated Tweets in our data. We used seven available lists of bot accounts (used in \cite{broniatowski2018weaponized}) to identify bot accounts. These lists \cite{LeePolluters,cresci2017paradigm,varol2017online,FrommerTwitterList,cresci2015fame,cresci2018social,nbcrusstroll} which id's of 8076 bots generally overlap in time with our study period (except for \cite{LeePolluters} in which the data was collected from December 30, 2009 to August 2, 2010, thus possibly very few users from this list would be relevant to our data). We assessed i) the total proportion of bot accounts in our data, and those responsible for discrimination content, and ii) the spatial distribution of those accounts.

\section{Results}
\subsection{Classification}
\subsubsection{Training Data Representation} We found that the top 20 features that were most predictive of a Tweet being classified as containing discrimination occurred in more than 25\% of the discrimination Tweets made in the top 11 overall hate crime cities. We studied the spatial and temporal distribution of these top 20 features to assess the spatial (and temporal) balance of the training data. We first assessed the temporal changes in the use of features and found there was no significant difference between the \% use of any of the top features in each city between all pairs of consecutive years (two-tailed Student's t-test, $p>$0.05). We then compared the distribution of these top 20 features in the 11 high overall hate crime cities with the distribution in the remaining 89 cities (medium and low hate crime) and high correlation between the use of features in the two groups determined using the Spearman’s rank correlation ($\rho$=0.87, $p<$0.05). We also performed the same analysis for the entire keyword list, and found no significant difference between the 2 sets of cities in any consecutive years ($p>$0.05) and high spatial correlation ($\rho$=0.84, $p<$0.05). Therefore, as the overall feature and keyword distributions were similar spatially and over consecutive years, we found it sufficient to use the training data generated by the top 11 hate crime cities to classify Tweets made in the additional 89 cities.

\begin{figure}[t]
\centering
\includegraphics[width=\columnwidth]{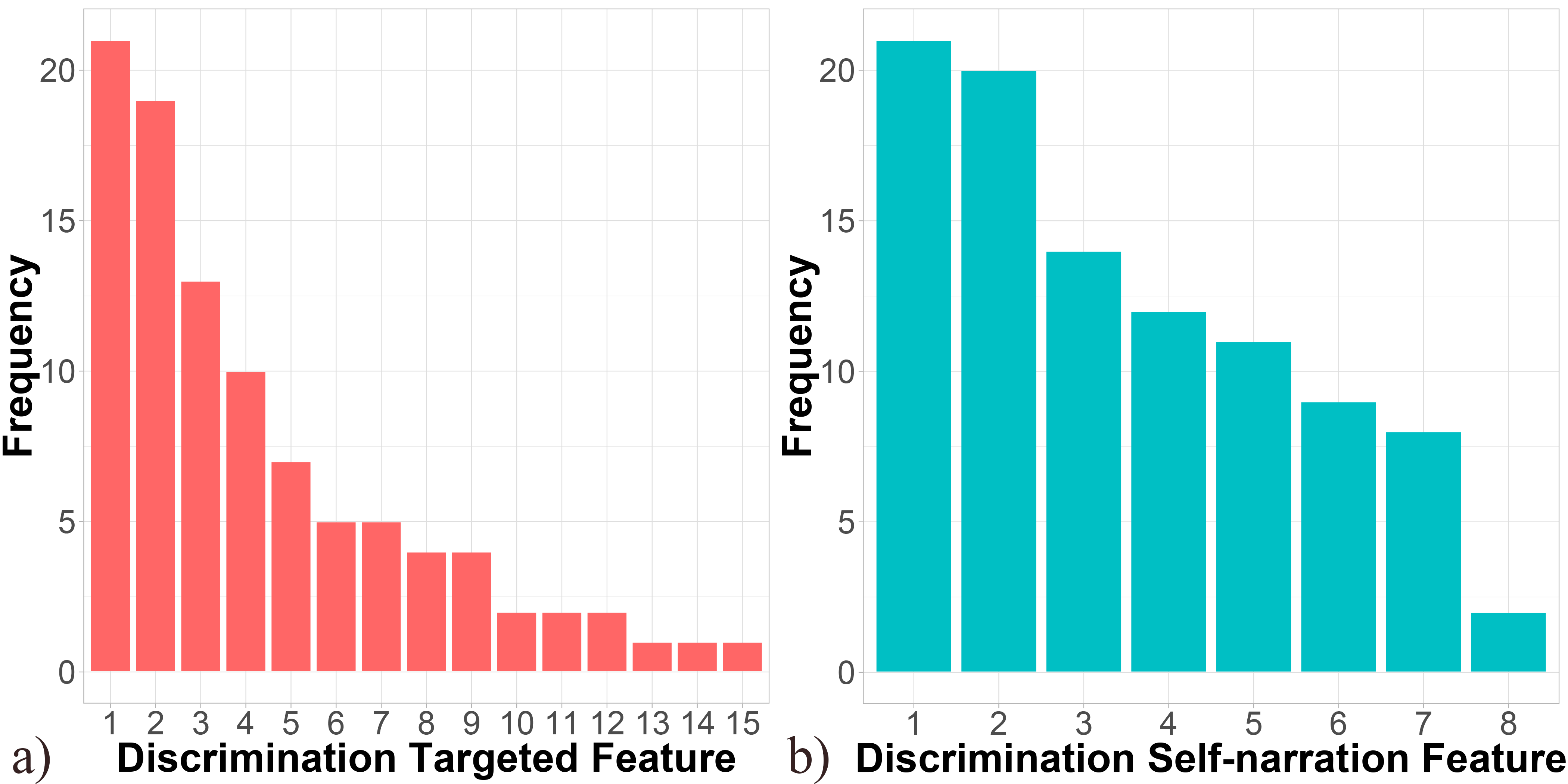}
\caption{Distribution of a) targeted and b) self-narration discrimination features (ranked) from the top 10 features in each of the 21 cities with the highest number of race, ethnicity or national-origin based hate crimes.}
\label{fig:feat_dist}
\end{figure}

\subsection{Spatial Distribution}
\paragraph{Crime distribution}
Phoenix, AZ had the highest number of race, ethnicity or national-origin based hate crimes over the 6 years (566). Conversely, Castleton, VT and Riverton, WY had 0 race, ethnicity or national-origin based hate crimes, and 26 cities had 9 total or less. The biggest differential in rank of race, ethnicity or national-origin based hate crimes and total discrimination Tweets was in Castleton, VT (lowest number of race, ethnicity or national-origin based hate crimes and 17th highest proportion of discrimination Tweets), and Montpelier, ID (in the 19th lowest by number of race, ethnicity or national-origin based hate crimes, and 5th highest proportion of discrimination Tweets).

\begin{figure}[t]
\centering
\includegraphics[width=\columnwidth]{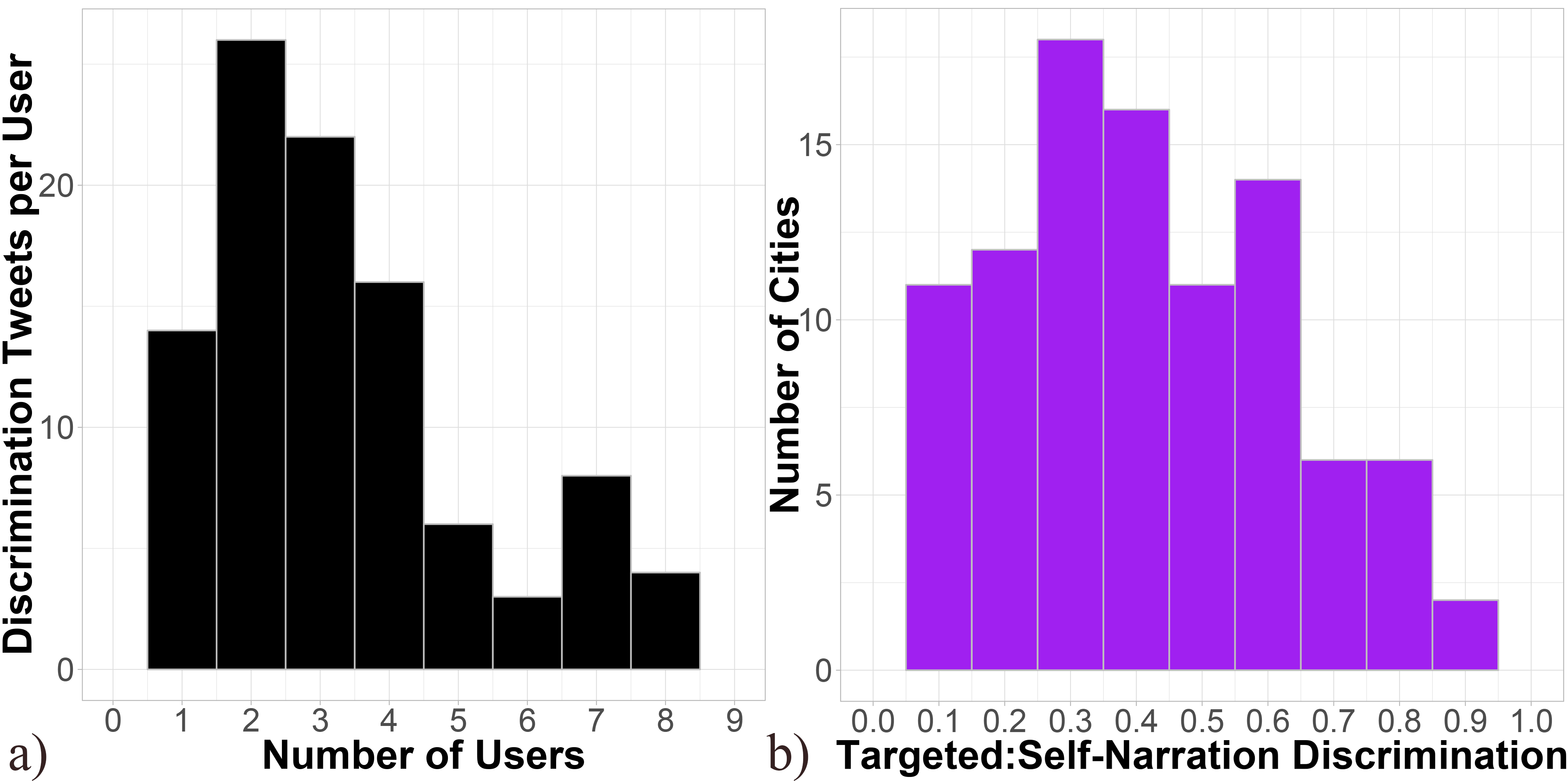}
\caption{Distribution of a) number of discrimination Tweets per user and b) ratio of targeted to self-narration discrimination in Tweets by city.}
\label{fig:users_dist}
\end{figure}

\paragraph{Social Media Feature Distribution}
Overall we found that most of the top features identified were used consistently across cities, but notably, there were some features that were found specifically in particular cities. For studying the top discrimination feature distribution we used the top 10 features in each of the top 21 cities ranked based on race, ethnicity or national-origin based Hate crimes (the 21st was San Francisco, which added geographic diversity, also San Francisco ranked highly for hate crimes in general (13th)). Figure \ref{fig:feat_dist} illustrates the distribution of the top discrimination features across a) targeted and b) self-narration Tweets. In terms of top features, there were three used in all of the top 21 cities ranked by race, ethnicity or national-origin based hate crimes: \emph{f*cking n*ggers} (* added to censor offensive language) which was the top feature in 10 of the top discrimination hate crime cities, \emph{most racist person} which appears in all top 21 cities (and is the top 1 or 2 feature in 9 cities), \emph{white trash} which is the top 1 or 2 feature in 4 cities, Indianpolis, IN, Cincinnati, OH, Las Vegas, NV and San Diego, CA. Three features were only found in single cities: \emph{f*cking wiggers} (Seattle, WA), \emph{insane redskin trash} (Washington, DC), \emph{n*gger is like} (Seattle, WA). Consistency of some of the top features, but appearance of some features only in specific places indicate there is some geographic variation that may not have been discovered if we did not ensure the training data was well distributed spatially. The most common feature was used in targeted discrimination and 15 of the 23 top features were used in targeted discrimination Tweets and the rest in self-narration discrimination Tweets, except \emph{white trash} which we found is used in both contexts (Figure \ref{fig:feat_dist}).

In 36 of the 100 cities, the proportion of discrimination that is targeted was higher than that of self-narration of discrimination experiences. The top and bottom ranked cities for targeted to self-narration discrimination ratio were: Woodbury, NJ (0.91), Greeneville, TN (0.89), Norman, OK (0.82), Missoula, MT (0.8), Neptune, NJ (0.79) and West Jordan, UT (0.02), Providence, RI (0.02), Riverton, WY (0.02), Mandan, ND (0.04), San Antonio, TX (0.05).

\subsection{Descriptive Analyses of Discrimination}
The overall number of users by city who discuss any discrimination on Twitter has a long tail distribution, with a mean of 731 users and standard deviation of 1448. When examining the number of users in relation to the number of discrimination Tweets, the largest number of cities have users with 2-3 discrimination Tweets (Figure \ref{fig:users_dist}a). Cities with a relatively higher proportion of Tweets to users (more discrimination Tweets by each user on average) are: Miami Beach, FL (8.7), Neptune Township, NJ (8.1), Taunton, MA (8.1) and Omaha, NE (8.0). In examining the proportion of discrimination Tweets that are targeted, compared to self-narration of discrimination by city, gives a more consistent distribution (Figure \ref{fig:users_dist}b), with 9 cities having a ratio above 0.7. These cities are: Woodbury, NJ (0.91), Greeneville, TN (0.89), Norman, OK (0.82), Missoula, MT (0.80), Neptune Town, NJ (0.79), Devils Lake, ND (0.79), Fullerton, CA (0.78), Milwaukee, WI (0.77) and Montpelier, ID (0.73).

\begin{table}[t!]\centering
\small
\begin{tabular}{l l l l}
\toprule
\textbf{Variable} & $\beta$ & std. err & $p$\\ 
\midrule
\% targeted Tweets            &      -0.087  &  0.066   &   0.188   \\
\% self narration Tweets     &       0.108 & 0.071  &  0.129  \\
Targeted:self narration  & 3.431   &   0.782  &      $<$0.001\textsuperscript{***} \\
\% white            &  0.030  &    0.028   &      0.280   \\
\% black            &     0.050          &  0.029  &  0.082.   \\
\% Asian     &   0.037           & 0.041   & 0.371\\
\% hispanic/latino   & 0.011 & 0.013   & 0.409   \\
\% foreign born   & 0.037 & 0.021             &   0.079.   \\
Median income    &  3.12e-6 & 1.13e-5&      0.783   \\
Population density    & 5.77e-5 & 3.70e-5 &     0.119   \\
\% female    & -0.098 & 0.125             &   0.434  \\
\% ages 18-64    & -0.035 & 0.026             &    0.178  \\
Intercept            &   5.330 & 6.947             &      0.443   \\
\midrule
\bottomrule
\addlinespace[1ex]
\multicolumn{4}{l}{
\textsuperscript{***}$p<0.001$, 
\textsuperscript{**}$p<0.01$, 
  \textsuperscript{*}$p<0.05$, 
  . $p<0.1$}
\end{tabular}
\caption{Regression results (all social media and other covariates predicting hate crimes). Full model.}
\end{table}

\begin{table}[t!]\centering
\small
\begin{tabular}{l l l l}
\toprule
\textbf{Variable} & $\beta$ & std. err & $p$\\ 
\midrule
Targeted:self narration  & 2.668   &   0.485 &   3.82e-08\textsuperscript{***} \\
\% black            &   0.018         &  0.006  &  0.0043\textsuperscript{**}   \\
\% foreign born   & 0.059 & 0.010            &   $<$ 0.001\textsuperscript{***}   \\
Intercept            &  1.620 & 0.302             &    $<$ 0.001\textsuperscript{***}   \\
\midrule
\bottomrule
\addlinespace[1ex]
\end{tabular}
\caption{Regression results (all social media and other covariates predicting hate crimes). Stepwise model.}
\end{table}

\begin{table}[b!]\centering
\small
\begin{tabular}{l r r l}
\toprule
\textbf{Variable} & $\beta$ & std. err & $p$\\ 
\midrule
\% targeted Tweets            &  -0.037  &  0.057   &  0.508  \\
\% self narration Tweets     &  0.118 & 0.061 & 0.053.  \\
Targeted:self narration  & 1.934  & 0.686  &      0.005\textsuperscript{**} \\
\% white            &  0.040  &  0.024  &      0.101  \\
\% black            &    0.056    &  0.025 &  0.026\textsuperscript{*}   \\
\% Asian     &  0.064   & 0.036  & 0.072.\\
\% hispanic/latino   & 0.015 & 0.011  & 0.185  \\
\% foreign born   & 0.002 & 0.019   &   0.926   \\
Median income    &  9.08e-7 & 9.83e-6 &      0.926   \\
Population density    & 5.77e-5 & 4.0e-5 &  0.154  \\
\% female    & -0.031 & 0.109   &  0.776 \\
\% ages 18-64    & 0.024 & 0.023   &  0.302 \\
Intercept    & -2.080 & 6.030   &  0.730  \\
\midrule
\bottomrule
\addlinespace[1ex]
\multicolumn{4}{l}{
\textsuperscript{***}$p<0.001$, 
\textsuperscript{**}$p<0.01$, 
  \textsuperscript{*}$p<0.05$, 
  . $p<0.1$}
\end{tabular}
\caption{Regression results (all social media and other covariates predicting hate crimes). Full model. Outliers removed.}
\end{table}

\begin{table}[b!]\centering
\small
\begin{tabular}{l l l l}
\toprule
\textbf{Variable} & $\beta$ & std. err & $p$\\ 
\midrule
Targeted:self narration  & 1.315 &   0.445 &   0.003\textsuperscript{**} \\
\% self narration Tweets     & 0.081 & 0.051  & 0.112  \\
\% black            &   0.013        &  0.005  &  0.015\textsuperscript{*}   \\
Population density   & 8.81e-5 & 3.06e-5            &   0.004\textsuperscript{**}   \\
Intercept      &  2.157 & 0.3002   &    6.66e-13\textsuperscript{***}   \\
\midrule
\bottomrule
\addlinespace[1ex]
\multicolumn{4}{l}{
\textsuperscript{***}$p<0.001$, 
\textsuperscript{**}$p<0.01$, 
  \textsuperscript{*}$p<0.05$, 
  . $p<0.1$}
\end{tabular}
\caption{Regression results (all social media and other covariates predicting hate crimes). Stepwise model. Outliers removed.}
\end{table}

\subsection{Hate Crimes and Social Media Relationship}
Upon visual inspection, we noticed that a few cities had a very high number of hate crimes (in general, including specifically race, ethnicity or national-origin based hate crimes): Phoenix, AZ, Boston, MA, Columbus, OH, Los Angeles, CA, New York, NY, Seattle, WA and Kansas City, MO. These cities are known for various reasons to have high number of hate crimes. Phoenix has a bias crimes unit which most major cities do not have, and Phoenix police say they look to be as thorough as possible when it comes to hate crimes, in contrast with other cities which take a more relaxed approach to personal attacks \cite{CrenshawPhoenix}. Seattle also records precinct-level hate crime data (unlike other cities except NYC). New York City and Los Angeles are the largest metropolitans in the United States, and Boston (and the state of Massachusetts) has more agencies contributing information to the FBI than most other states \cite{JarmanningMA}. The higher number of reported hate crime incidences in Columbus, OH has been recognized and attributed to it's disclusion of heightened punishments for crimes such as assault or murder though, and lack of inclusion for protections for sexual orientation, gender identity, age, disability, or military status \cite{KocutColumbus}. Finally, Kansas City, MO is also known to be one of the top crime (in general) cities in the country \cite{AlcockMO}. We thus performed the regression analysis both with and without these outlier cities. 

Regression results show that the ratio of discrimination Tweets that are targeted to self-narrations has a positive relationship ($\beta>0$) with the number of race, ethnicity or national-origin based hate crimes in a city, and this variable is significant when controlling for all of the demographic and other city attributes (Tables 1-4). The stepwise model shows that percentage black and foreign born also contribute a significant portion of the explained variance. When outliers are removed, the main difference in model results is that the stepwise model shows the total percent of discrimination Tweets that are self-narration , percent black and population density of cities to also contribute to explained variance along with the targeted to self-narration ratio (all with positive coefficients).

\begin{table}[t!]\centering
\small
\begin{tabular}{c c}
\toprule
\textbf{Feature} &  Times more likely in discrimination text \\ 
\midrule
crime    &  7.2    \\
fear    &      7.8      \\
hearing    &       8.5         \\
dominant personality &  8.4    \\
sadness            &   9.7   \\
anonymity     &   12.8 \\
ugliness            &   14.0   \\
work            &   15.3   \\
neglect            &    38.6   \\
terrorism            &   39.0   \\
\midrule
\bottomrule
\end{tabular}
\caption{EMPATH features more common in discrimination versus non-discrimination Tweets.}
\label{tab:empathratio}
\end{table}


\subsection{Linguistic Results}
\subsubsection{City-level Results} Of the EMPATH features selected based on their potential relation to our outcome, in discrimination Tweets, both positive and negative emotion were significant predictors of hate crimes. Surprisingly, positive emotion had a positive significant relationship while negative emotion a negative relationship with the number of race, ethnicity or national-origin based hate crimes. Disappointment, money, and night all had a positive, significant relationship with the number of hate crimes. Work-related features had a negative significant relationship.

\subsubsection{User-level Results} In assessment of the linguistic characteristics of users who discuss relatively high amounts of discrimination ($>$21 discrimination Tweets), we found that correlation of the resulting EMPATH categories in their discrimination Tweets was very high both with the discrimination Tweets of users who made only 1 discrimination Tweet, as well as with those who made between 1 and 21 ($\rho$ = 0.99, $p<$0.05, for both correlations). This indicates that linguistic characteristics are consistent amongst those who post a lot of discrimination versus a little. Pearson correlation between the linguistic characteristics common to non-discrimination and discrimination Tweets was fairly high but not as similar ($\rho$ = 0.80, $p<$0.05). \subsubsection{Tweet-level Results} We further examined the specific EMPATH characteristics that did not fall in line with the correlation via the normalized mean of each feature count. Table \ref{tab:empathratio} shows the top 10 features with a normalized mean in discrimination tweets compared to the normalized mean for the same feature in non-discrimination Tweets. In general, negative features that do all have some intuitive relation with discrimination are most common. The average normalized mean ratio is 3.6; and so these features are at least twice or more times likely to be in discrimination versus non-discrimination text (minimum is 7.2 times in the table).

\begin{table}[htbp]\centering
\small
\begin{tabular}{l r r l}
\toprule
\textbf{Variable} & $\beta$ & std. err & $p$\\ 
\midrule
positive emotion            &      1.633  &   0.568   &     0.004\textsuperscript{**}   \\
negative emotion    &       -1.241        &    0.624  &       0.047\textsuperscript{*}  \\
disappointment            &  0.471  &  0.219  &   0.032\textsuperscript{*}   \\
work            &  -1.660         &    0.561& 0.003\textsuperscript{**}  \\
money     & 0.497     & 0.252             &  0.049\textsuperscript{*}\\
night            &  0.467 & 0.156             &      0.003\textsuperscript{**}   \\
Intercept    &    3.163  &    0.113  &      \textless 0.001\textsuperscript{***} \\
\midrule
\bottomrule
\addlinespace[1ex]
\multicolumn{4}{l}{
\textsuperscript{***}$p<0.001$, 
\textsuperscript{**}$p<0.01$, 
  \textsuperscript{*}$p<0.05$, 
  . $p<0.1$}
\end{tabular}
\caption{Empath regression results, predicting race, ethnicity or national-origin discrimination-motivated hate crimes. Stepwise model.}
\end{table}

\subsection{Discrimination Content by Bots}
We found a minimum of zero\% (18 cities) to maximum of 31.6\% (Washington, D.C.) discrimination users that were classified as bots based on the lists of known bots, with a mean of 7.8\% (sd: 6\%) (fairly spatially consistently distributed, with only 11 cities above 15\%). Some example Tweets by identified bots are \emph{``Giants playing terribly against a terrible franchise. Enjoy gloating skins fans, you're still white trash. \#Giants \#redskins''} (New York, NY). \emph{``@Usernameredacted it a proven fact that \#blackpeople are the most racist people out there''} (San Francisco, CA). Overall there were many themes in the bot posts that were classified as indicating discrimination. We decided to keep Tweets from the bots in our analysis, as these Tweets would be visible to followers as they were posted, though we remark that these should be further investigated or noted in any further analyses of the causal reasons for or implications of discrimination in social media.

\section{Discussion}
\textbf{Summary of Contributions to Social Media Research} In this work, we study the characteristics of race, ethnicity and national-origin based discrimination on social media spatially, as well as hate crimes motivated by these biases across the United States. In creating the spatially diverse training data set of social media discrimination, we found that most of the features predictive of discrimination were common, but there are examples of less common features that only appear in select cities. As well, we showed that there is a larger distribution of features in discrimination Tweets that are \emph{targeted} compared to those that are \emph{self-narration} of discrimination in the cities with the most race, ethnicity and national-origin based hate crimes. In terms of the relationship between social media discrimination and race, ethnicity and national-origin based hate crimes, the proportion of social media discrimination that is targeted was significantly related to the number of hate crimes. When not considering specific cities with outlier numbers of crimes, the proportion of social media discrimination that is self-narration was also significant. Linguistically, we identified features more common in discrimination Tweets versus non-discrimination Tweets, and also showed that positive and negative emotion, as well as disappointment, money, night and work were significantly related to race, ethnicity  or national-origin based hate crimes by city. The surprising significance of positive emotion may be potentially related to high-levels of emotion in general in discrimination, or positivity in response to self-narration of discrimination experiences \cite{tynes2012online}. The ubiquity of emotion in discrimination is also supported by the increased frequency of the empath feature \emph{sadness} in discrimination Tweets (Table \ref{tab:empathratio}). Finally, we also showed that there was race-based discrimination from existing, recognized lists of Twitter bots, linked to most of the 100 cities considered, and most predominantely in Washington, D.C..

\textbf{Implications of Analysis} We stress that while our work makes no causal claims directly between discussion on social media and crimes, findings from this study are pertinent for better discrimination surveillance and mitigation efforts. In particular, as this work shows that social media may significantly explain some of the variation in hate crimes, discrimination on social media should be studied further to understand, contrast and assess the possible synergies of online and physical world discrimination. The linguistic analysis highlights the opportunities of social media to dissect and understand more about different types of discrimination in the country. These opportunities are discussed further in the Future Work section below. As there has been some concern regarding the criteria and consistency in how hate crimes are reported in different cities, social media also provides a different measure of systemic discrimination by which to augment our understanding of this phenomenon; for example discrimination on social media encompasses that which doesn't necessary elevate to the level of a crime or for which there are no laws mandating reporting in a particular region (e.g. sexual-orientation based discrimination), or location (e.g. sub-city level regions), but such day-to-day negative insults can be internalized, still impact communities and should be ascertained \cite{williams2003racial}. 

\textbf{Future Work} 
Beyond this spatial analysis, and given newly released statistics from the FBI that show a 17\% increase in hate crimes nationwide, in 2017 \cite{VOAcrimesrise}, a temporal analysis of both discrimination on social media and hate crimes could be of relevance. It should be noted that changes in Twitter (or any company's) policies around hate speech should be carefully considered if attempting to unpack temporal changes or causal mechanisms. In generating the spatially balanced training data, we did find a sudden drop in the number of Tweets containing discrimination keywords across all cities in 2016 as compared to previous years. This drop was likely caused by Twitter's strategy in decreasing hate speech (e.g. using e-mail and phone verification) announced in December 2015 \cite{TwitterChange}. This change coupled with the aggregation of hate crimes motivated by biases in different ways across the years would have to be accounted for in any temporal analysis or assessment of discrimination based on more specific biases. Though the drop based on Twitter's actions was spatially consistent across all Tweets, and therefore not a concern in the context of our spatial analysis, it is the type of mitigation that this work can potentially inform (via the types of features or the need for spatially-different linguistic features and differences to be considered). Further, our finding that the proportion of race, ethnicity or national-origin discrimination online that is \emph{targeted} is a significant predictor in the regression model indicates a focus on this measure. Further analyses should also consider unpacking differences in the relative prevalence of hate crimes and social media discrimination in different cities, discrimination against different groups based on the language/terms used, incorporation of non-english languages and communication through emojis \cite{barbieri2016cosmopolitan}.


\subsubsection{Acknowledgments.}
Blank for now.


\bibliography{ICWSM_socmedhat}
\bibliographystyle{aaai}
\end{document}